\documentclass[onecollarge,natbib]{svjour2}
\bibpunct{[}{]}{;}{n}{}{,} 
\smartqed  
\usepackage{graphicx}
\journalname{Archive of Applied Mechanics}
\def\be{\begin{equation}} \def\ee{\end{equation}} \def\bea{\begin{eqnarray}}
\def\eea{\end{eqnarray}} \def\nnb{\nonumber}

\begin{document}

\title{
The Born series for $S$-wave quartet $nd$ scattering 
at small cutoff values 
}

\author{
Shung-Ichi Ando 
}

\institute{Shung-Ichi Ando \at
Department of Physics Education, 
Daegu University,
Gyeongsan 712-714,
Republic of Korea \\
              Tel.: +82-53-8506975\\
              Fax: +82-53-8506979\\
              \email{sando@daegu.ac.kr}           
}

\date{Received: date / Accepted: date}

\maketitle

\begin{abstract}
Perturbative expansions, the Born series, of the scattering length
and the amplitude of $S$-wave neutron-deuteron scattering for spin quartet
channel below deuteron breakup threshold are studied in pionless effective 
field theory at small cutoff values. A three-body contact interaction
is introduced when the integral equation is solved using the small cutoffs. 
After renormalizing the three-body interaction by using the scattering
length, we expand the integral equation as the 
ordinary and inverted Born series. 
We find that the scattering length and the phase shift are considerably
well reproduced with a few terms of the inverted Born series 
at relatively large cutoff values, $\Lambda\simeq 100$~MeV.
\end{abstract}

\section{Introduction}
\label{intro}

After Weinberg suggested the application
of chiral perturbation theory,
a low energy effective field theory of QCD, to nuclear force~\cite{w}, 
a lot of works have been done during last two decades.
For reviews, see, e.g., Refs.~\cite{bk-arnps02,bh-pr06,ehm-rmp09,me-pr11}.
Meanwhile, 
new facilities for the rare isotope beams, 
in which exotic nuclei near the neutron (proton) drip line
can be created,
provide us a new opportunity for 
investigation~\cite{j-pr04,zetal-pr93,t-jpgnpp96}. 
Such a limit close to the drip line 
where the single neutron (proton) separation energy vanishes 
can be theoretically recognized as unitary limit (to be discussed below),
and thus the unitary limit would be a good theoretical starting point
to study the exotic nuclei near the nuclear drip line.
In this work, we discuss a perturbative expansion
of an integral equation, as the inverted Born series
which can be related to the unitary limit,
by sending the cutoff parameter to small values
(along with the ordinary Born series).
We study it in $S$-wave neutron-deuteron ($nd$) scattering 
for spin quartet channel in pionless effective field theory (EFT). 

Pionless EFT~\cite{crs-npa99,ah-prc05} 
is one of low energy EFTs for few nucleon systems,
in which one chooses a typical momentum scale $Q$ 
smaller than the pion mass, $m_\pi$, and thus the pions are 
regarded as heavy degrees of freedom and integrated out of 
effective Lagrangian. Thus the large scale of the pionless EFT
is $\Lambda \simeq m_\pi$, and the theory provides us a systematic
expansion scheme in terms of $Q/\Lambda$.
In applications of the pionless EFT, much more attention has been 
paid to $S$-wave $nd$ scattering for 
spin doublet $(S=1/2)$ channel because the one-nucleon-exchange  
interaction becomes ``singular''~\cite{bhk-prl99,bhk-npa00}. 
To control the singularity one promotes a three-body contact 
interaction to leading order (LO), and the coupling
constant of the contact interaction exhibits so called limit-cycle
when the scale parameter $\Lambda$ is sent to the asymptotic limit.  
This behavior is understood to be associated with ``Efimov effect''.
On the other hand, an application of pionless EFT 
to $S$-wave $nd$ scattering for spin quartet ($S=3/2$) channel 
is regarded well known
because the phase shift $\delta_0$ of the quartet $nd$ scattering 
is almost perfectly described by two effective range parameters 
in two-body deuteron channel, and no cutoff dependence is 
reported~\cite{bvk-plb98,bhvk-prc98}.

In this work, we revisit the $S$-wave spin quartet $nd$ scattering 
below deuteron breakup threshold
in pionless EFT and reduce the cutoff parameter $\Lambda$
smaller than $m_\pi$. The small scale 
limit might be interesting for studying, e.g., 
a relation between the pionless EFT 
and a Halo/Cluster EFT~\cite{bhk-npa02} in which one may choose 
a typical scale smaller than the deuteron breakup momentum, 
and the deuteron is never broken into two nucleons 
and could be regarded as a cluster or an elementary field~\cite{sia}. 

When the value of $\Lambda$ is sent to significantly smaller 
than $m_\pi$, the cutoff dependence emerges even in the quartet channel.
Thus we introduce a three-body contact interaction 
when we solve the integral equation using the small cutoff values.
After renormalizing the strength of the three-body interaction 
by using the scattering length $a_4$ of the $S$-wave spin quartet $nd$ 
scattering we expand the integral equation 
for the scattering length $a_4$ and the scattering amplitude
in terms of the Born series up to next-to-next-to leading order 
(NNLO)~\footnote{
We note that this is not an expansion scheme in EFT, but 
in terms of the Born series. 
We do not have a clear expansion parameter as $Q/\Lambda$. 
}. 
We expand the Born series around so called trivial and non-trivial 
fixed point studied in the renormalization group 
analysis by Birse, McGovern, and Richardson~\cite{bmr-plb99}. 
The trivial fixed point corresponds to weak coupling limit
where all interactions vanish, 
whereas the non-trivial fixed point does to 
unitary limit where the scattering length becomes infinity 
or the binding energy vanishes.
Around the unitary limit, the Born series is 
realized as the inverted Born series and expanded
around the inverse of the amplitude~\cite{sia-12}.
We find that, by reducing the cutoff value,
both $a_4$ and $\delta_0$ are 
considerably well reproduced by a few terms of the inverted Born series
at relatively large cutoff values, $\Lambda\simeq 100$~MeV.

This work is organized as the following.
In Sec.~\ref{sec:1}, effective Lagrangian is displayed.
In Sec.~\ref{sec:2}, integral equations for the scattering length and 
the amplitude are given, and the coupling of the three-body interaction 
is renormalized by the scattering length $a_4$.
In Sec.~\ref{sec:3}, the scattering length and the amplitude are expanded
in terms of the ordinary and inverted Born series, 
and the numerical results are obtained.
Finally, in Sec.~\ref{sec:4}, the discussion and conclusions are presented. 

\section{Effective Lagrangian}
\label{sec:1}

The effective Lagrangian for the $S$-wave spin quartet $nd$ scattering
in pionless EFT reads~\cite{ah-prc05,bhk-npa00} 
\bea
{\cal L} &=& {\cal L}_N 
+{\cal L}_t
+{\cal L}_{3}\,,  
\eea
where ${\cal L}_N$ is the standard one-nucleon Lagrangian
in the heavy-baryon formalism,
\bea
{\cal L}_N &=&
N^\dagger \left\{
iv\cdot D 
+ \frac{1}{2m_N}\left[
(v\cdot D)^2 
-D^2\right]
+ \cdots
\right\}N \,,
\eea
where $v^\mu$ is a velocity vector satisfying a condition $v^2=1$, 
$D_\mu$ is the covariant derivative, and $m_N$ is the nucleon mass.
The dots denote higher order terms which are not relevant in the 
present work.
${\cal L}_t$ is dibaryon effective
Lagrangian for $^3S_1$ state,
\bea
{\cal L}_t &=&  
\sigma_t t_i^\dagger \left\{
iv\cdot D
+ \frac{1}{4m_N}\left[
(v\cdot D)^2 
-D^2 \right]
+ \Delta_t
\right\}t_i
-y_t \left\{
t_i^\dagger \left[ 
N^T P_i^{({}^3S_1)}N
\right] + h.c.
\right\}
+ \cdots \,,
\eea
where 
$\sigma_t$ is a sign factor, 
$\sigma_t = \pm 1$.
$t_i$ is a dibaryon field for spin triplet (${}^3S_1$) state.
$D_\mu$ is the covariant derivative for the dibaryon field.
$\Delta_t$ is the mass difference between 
the dibaryon and two nucleons,
$m_{t} = 2m_N + \Delta_{t}$. 
$y_t$ is a coupling constant for 
dibaryon-nucleon-nucleon ($dNN$) vertex.
$P^{({}^3S_1)}_i$ is a projection operator for 
two nucleons in the ${}^3S_1$ state, 
\bea
P^{({}^3S_1)}_i = \frac{1}{\sqrt8} \tau_2\sigma_2\sigma_i\,,
\eea
where $\tau_a$ and $\sigma_i$ are Pauli matrices for 
isospin and spin, respectively.
Those two constants, $\Delta_t$ and $y_t$, and the sign 
factor $\sigma_t$ ($=-1$) are determined by two effective range 
parameters, deuteron binding momentum $\gamma$ and 
effective range $\rho_d$, in  
$NN$ scattering for 
$^3S_1$ channel. 

We obtain ${\cal L}_3$, an effective Lagrangian 
of three-body contact interaction 
in terms of a nucleon and a dibaryon field
for the $S$-wave 
spin quartet $nd$ state, as
\bea
{\cal L}_3 &=& 
-\frac{m_N}{6}g(\Lambda)y_t^2 
N^\dagger \left[
(\vec{\sigma}\cdot\vec{t})^\dagger (\vec{\sigma}\cdot \vec{t})
+3(\vec{\sigma}\cdot\vec{t}) (\vec{\sigma}\cdot \vec{t})^\dagger
\right] N
+ \cdots \,.
\label{eq;L3}
\eea 
The expression of the interaction is the same as that of the spin
projection operator of spin-1/2 and spin-1 field into spin-3/2 state.
See, e.g., Eq.~(A4.37) in Ref.~\cite{textbook}.\footnote{
There are some variations of the spin 3/2 projection operator 
for the spin 1 and 1/2 fields:
\bea
P_{3/2}^{ij} = \frac16(\sigma^i\sigma^j + 3 \sigma^j\sigma^i)
= \frac23\delta^{ij} - \frac13i\epsilon^{ijk}\sigma^k
= \delta^{ij} -\frac13\sigma^i\sigma^j\,.
\nnb
\eea
}
We note that the three-body interaction is a higher order
term in the $S$-wave spin quartet $nd$ scattering in pionless EFT, 
and it is not necessary to be introduced when the value of the cutoff
$\Lambda$
is about 
$m_\pi$ or larger, $\Lambda \ge m_\pi$,
at leading order (LO). 
In addition, 
as we will see below,
it becomes ineffective when we solve the integral equation 
using the ordinary cutoff value. 
%
We introduce it, however, so as
to reproduce the scattering length $a_4$ 
in the quartet channel (and the strength
of the coupling $g(\Lambda)$ is adjusted) when the $\Lambda$ value 
is reduced significantly smaller than $m_\pi$, $\Lambda < m_\pi$.   
%

\section{Integral equations}
\label{sec:2}

\subsection{Integral equation of 
$S$-wave $nd$ scattering amplitude for quartet channel}
\label{sec:2.1}

\begin{figure}
\begin{center}
 \includegraphics[width=14cm]{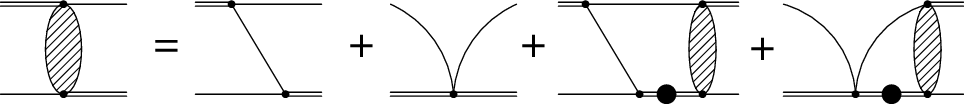}
\caption{Diagrams for $S$-wave $nd$ scattering 
in spin quartet ($S=3/2$) channel.
A shaded blob denotes 
scattering amplitude 
a line (a double-line) nucleon (dibaryon), and
a double line with a filled circle dressed dibaryon propagator
(see Fig.~\ref{fig;dibaryonpropagator}).}
\label{fig;IE-quartet}
\end{center}
\end{figure}

\begin{figure}
\begin{center}
\includegraphics[width=12cm]{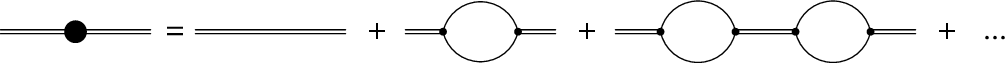}
\caption{Diagrams for a dressed dibaryon propagator.
See the caption of Fig.~\ref{fig;IE-quartet} as well.}
\label{fig;dibaryonpropagator}
\end{center}
\end{figure}

Diagrams for $S$-wave $nd$ scattering
for the spin quartet ($S=3/2$) channel 
are given in Fig.~\ref{fig;IE-quartet}, whereas
those for the two-body part in Fig.~\ref{fig;IE-quartet}, 
the dressed dibaryon propagator, are given 
in Fig.~\ref{fig;dibaryonpropagator}. 
The integral equation of the $S$-wave spin quartet $nd$ scattering 
without the three-body contact 
interaction is known as~\cite{bhvk-prc98}
\bea
t(p,k) &=& 
-\frac{8\pi}{m\rho_d} K^{(a)}(p,k;E)
\nnb \\ && 
- \frac{2}{\pi}\int^\Lambda_0dq
K^{(a)}(p,q;E)
\frac{q^2 t(q,k)}{
-\gamma
+\frac12\rho_d\left(
\gamma^2
-\frac34q^2
+mE
\right) 
+ \sqrt{\frac34q^2-mE}}\,,
\label{eq;t}
\eea
where $t(p,k)$ is the half-off-shell scattering amplitude
and $p$ ($k$) is the magnitude of off-shell final state 
(on-shell initial state) relative three momentum in CM frame.
$K^{(a)}(p,q;E)$ is the one-nucleon-exchange interaction,
\bea
K^{(a)}(p,q;E) = \frac{1}{2pq}\ln\left(
\frac{p^2+q^2+pq-mE}{p^2+q^2-pq-mE}
\right)\,,
\eea
where $E=-\frac{\gamma^2}{m_N} + \frac{3}{4m_N}k^2$: 
$\gamma$ and $\rho_d$ are 
the deuteron binding momentum 
and the effective range, respectively.
A sharp cutoff $\Lambda$ is introduced in the loop integral.

We note that because a singularity at $q\simeq 197$\,MeV 
which represents an unphysical deeply bound state 
appears in the propagator,
the dressed dibaryon propagator is usually expanded 
in terms of the effective range, $\rho_d$,
to avoid an effect from the unphysical bound state.
However, we are interested in reducing the cutoff value,
less than $\Lambda =197$~MeV, 
so we do not expand 
the dressed dibaryon propagator in the integral equation in 
Eq.~(\ref{eq;t}).

Thus the on-shell T-matrix is given by
\bea
T(k,k) = \sqrt{Z_d} t(k,k) \sqrt{Z_d}\,,
\label{eq;T}
\eea
where $Z_d$ is the deuteron wavefunction normalization factor,
$Z_d = \gamma\rho_d/(1-\gamma\rho_d)$. 
In terms of the half-off-shell scattering amplitude $T(p,k)$,
we have the integral equation as
\bea
T(p,k) &=& - \frac{8\pi}{m_N\rho_d}Z_d K^{(a)}(p,k;E) 
\nnb \\ &&
- \frac{8}{3\pi} \int^\Lambda_0 dq 
K^{(a)}(p,q;E) 
\frac{
\gamma
+\sqrt{\gamma^2
+ \frac34(q^2-k^2)}
}{
1
-\frac12\rho_d\left(
\gamma
+\sqrt{
\gamma^2
+\frac34(q^2-k^2)}
\right)
}
\frac{q^2T(q,k)}{q^2-k^2-i\epsilon}\,.
\eea

\subsection{Integral equation for scattering length}
\label{sec:2.2}

The scattering length, $a_4$, of the 
spin quartet $nd$ scattering is obtained by 
the on-shell scattering amplitude with 
zero momentum as
\bea
T(0,0) = -\frac{3\pi}{m_N} a_4 \,.
\eea
Now we introduce a half off-shell amplitude
(or a half off-shell scattering length) as
\bea
T(p,0) = - \frac{3\pi}{m_N}a(p,0)\,,
\eea
where $a(0,0) = a_4$.
Thus we have an integral equation in terms of $a(p,0)$ as
\bea
a(p,0;\Lambda) &=& \frac{8Z_d}{3\rho_d} K^{(a)}(p,0;-B_2)
\nnb \\ && 
-\frac{8}{3\pi}
\int^\Lambda_0 dq
K^{(a)}(p,q;-B_2) 
\frac{\gamma+\sqrt{\gamma^2+\frac34q^2}}{
1-\frac12\rho_d\left(
\gamma
+ \sqrt{\gamma^2+\frac34q^2}
\right)} a(q,0;\Lambda)\,,
\label{eq;a-hosh}
\eea
where we have included the $\Lambda$ dependence in the 
scattering length $a(p,0;\Lambda)$ in the above expression
though $a(p,0;\Lambda)$ is insensitive to $\Lambda$ 
when $\Lambda\simeq m_\pi$ or larger.
$B_2$ is the deuteron binding energy, $B_2=\gamma^2/m_N$.

Now we choose $\Lambda = 197$\,MeV, and solve Eq.~(\ref{eq;a-hosh})
numerically. We have two input parameters, $\gamma$ and $\rho_d$,
whose values are known as 
$\gamma = 45.7\,$MeV and $\rho_d = 1.76\,$fm, 
and thus have
\bea
a_4^{th}\equiv a(0,0;\Lambda=197~\mbox{\rm MeV}) =  6.34\, {\rm fm}\,.
\label{eq;a4th}
\eea
We reproduce results obtained 
by Bedaque, Hammer, and van Kolck~\cite{bvk-plb98,bhvk-prc98},
$a_4=6.33\pm0.10\,$fm,
and Griesshammer~\cite{g-npa04},
$6.354\pm 0.020\,$fm, up to NNLO
in pionless EFT.
It agrees well with those obtained by 
Bedaque and Griesshammer~\cite{bg-npa00},
$6.8\pm 0.7\,$fm, up to NLO in EFT with perturbative pions,
H. Witala {\it et al.}~\cite{wetal-prc03}, 
6.321 ... 6.347~fm, from various 
combinations of modern $NN$ potentials and three-body forces,
and the experimental datum~\cite{dkn-plb71},
\bea
a_4^{exp.} = 6.35\pm 0.02\, {\rm fm}\,.
\eea

\begin{figure}[t]
\begin{center}
\includegraphics[width=7cm]{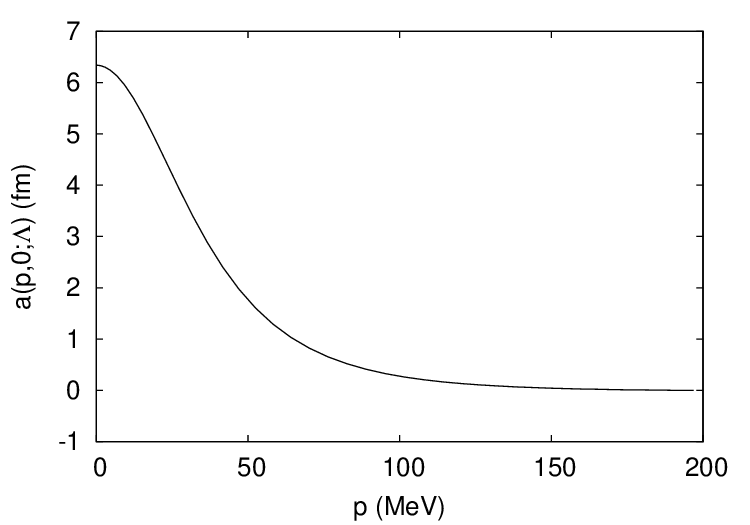}
\includegraphics[width=7cm]{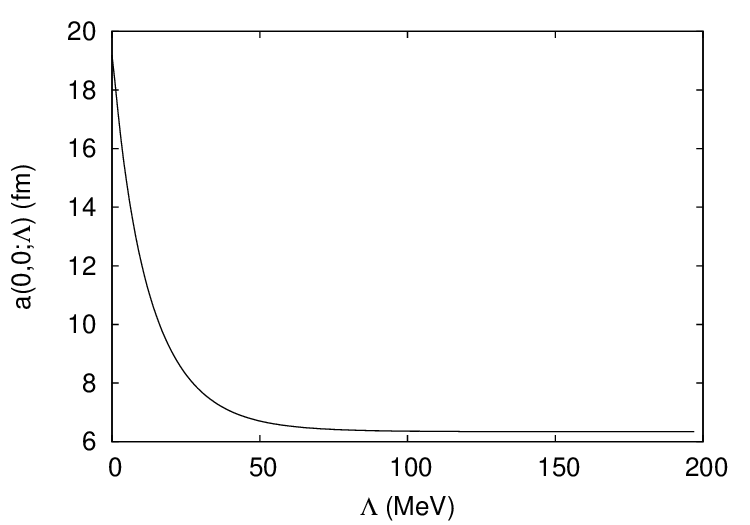}
\caption{
Left panel: Half off-shell $a(p,0;\Lambda)$ (fm) with $\Lambda=197\,$MeV
as a function of off-shell momentum $p$ (MeV).
Right panel: 
Cutoff dependence of scattering length $a(0,0;\Lambda)$ (fm) 
without the three-body interaction.
}
\label{fig;hoshaNLO}
\end{center}
\end{figure}
In the left panel of Fig.~\ref{fig;hoshaNLO},
we numerically calculate and plot 
the half off-shell scattering length, 
$a(p,0;\Lambda)$, at $\Lambda=197$~MeV as a function of the off-shell 
momentum $p$. 
We find that the off-diagonal part of the 
scattering length quickly decreases 
and becomes almost negligible at $p > m_\pi$. 
This fact may indicate the insensitivity of the amplitude for the quartet
channel to the value of $\Lambda$ when $\Lambda = m_\pi$ or larger,
$\Lambda\ge m_\pi$. 

We now reduce the value of $\Lambda$. 
In the limit where $\Lambda$ is sent to zero, we have the 
on-shell scattering length $a(0,0;\Lambda)$ as 
\bea
\lim_{\Lambda\to 0} a(0,0;\Lambda) 
= \frac83\frac{1}{\gamma(1-\gamma\rho_d)}
\simeq 19.19\, {\rm fm}\,.
\eea
In the right panel of Fig.~\ref{fig;hoshaNLO},
we plot the cutoff $\Lambda$ dependence 
of the scattering length $a(0,0;\Lambda)$.
One can see that there is almost no $\Lambda$ dependence
at $\Lambda>100$~MeV, whereas 
the significant cutoff dependence appears in the 
calculated scattering length $a(0,0;\Lambda)$
when the value of the cutoff is reduced less than 
$\gamma$ $(=45.7~\mbox{\rm MeV})$.

\subsection{Integral equation with three-body contact interaction}
\label{sec:2.3}

At the small cutoff region, especially being smaller than $\gamma$,
the scattering length $a(0,0;\Lambda)$ has the significant cutoff dependence.
To make the result cutoff independent,
as discussed before, 
we introduce the three-body contact interaction $g(\Lambda)$ 
which is supposed to take account 
of physics integrated out due to the small cutoff, 
and thus have the integral equation as
\bea
a(p,0) &=& \frac{8Z_d}{3\rho_d}
\left[
\frac{1}{\gamma^2+p^2}
+g(\Lambda)
\right]
\nnb \\ && - \frac{8}{3\pi}\int^\Lambda_0 dq \left[
K^{(a)}(p,q;-B_2) + g(\Lambda)
\right]
\frac{\gamma+\sqrt{\gamma^2+\frac34q^2}}{
1-\frac12\rho_d\left(
\gamma
+\sqrt{\gamma^2+\frac34q^2}
\right)
}a(q,0)\,,
\label{eq;a}
\eea
where we have removed the $\Lambda$ dependence 
from the half off-shell scattering length $a(p,0)$
because the $\Lambda$ dependence from the integral should
be cancelled by $g(\Lambda)$.

In the limit that the cutoff $\Lambda$ becomes large,
at $\Lambda=197$~MeV,
the coupling constant $g(\Lambda)$ should vanish,
$g(\Lambda) \to 0$. 
On the other hand, in the limit that the cutoff $\Lambda$ vanishes, 
one has
\bea
a_4^{th} = \frac83\frac{Z_d}{\rho_d}\left(
\frac{1}{\gamma^2} + g(0)
\right)\,,
\eea
where $g(0)=-12.56\cdots\,$fm$^2$ to reproduce the value of 
$a(0,0;\Lambda=197~\mbox{\rm MeV})$ in Eq.~(\ref{eq;a4th}).
Thus we consider a range of $g(\Lambda)$, 
$-12.56 \le g(\Lambda) \le 0$~(fm$^2$), in this work. 

\begin{figure}[t]
\begin{center}
\includegraphics[width=7cm]{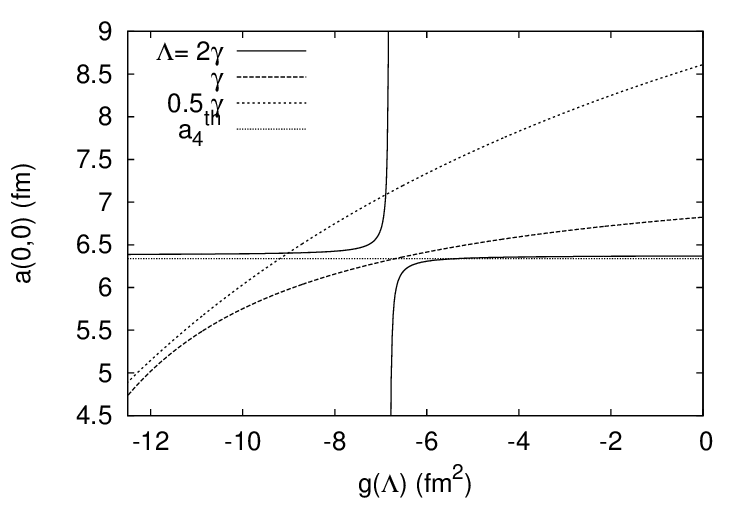}
\includegraphics[width=7cm]{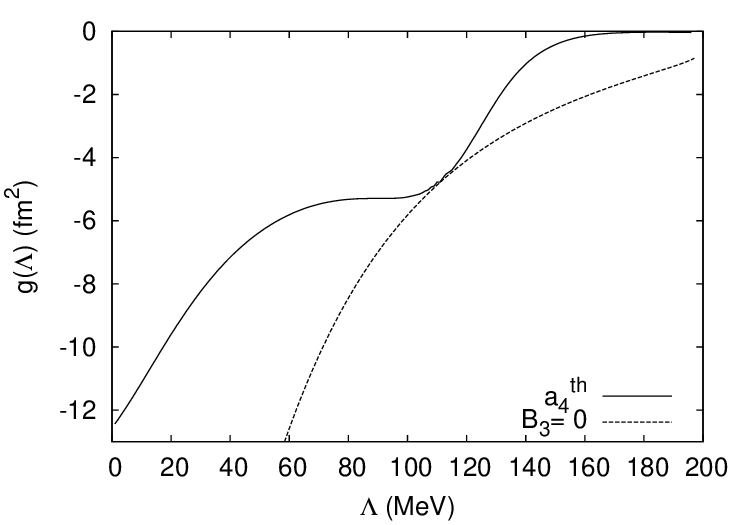}
\caption{
Left panel: $a(0,0)$ (fm)
in the cases of three fixed cutoff values $\Lambda = 2\gamma$, 
$\gamma$, and $0.5\gamma$ 
as functions of $g(\Lambda)$ (fm$^2$). 
A dashed horizontal line for $a_4^{th}$ is also included.
Right panel: $g(\Lambda)$ (fm$^2$) 
which reproduce $a_4^{th}$ (curve) and three-body 
bound state with $B_3=0$ (dashed curve) 
as functions of $\Lambda$ (MeV).  
}
\label{fig;a4Lam30}
\end{center}
\end{figure}

In the left panel of Fig.~\ref{fig;a4Lam30}, 
we numerically calculate $a(0,0)$ from Eq.~(\ref{eq;a}) and plot 
curves of $a(0,0)$ 
with three fixed values
of $\Lambda$, $\Lambda = 2\gamma$, $\gamma$, and 0.5$\gamma$
as functions of $g(\Lambda)$. 
A dashed horizontal line for $a_4^{th}$ is also included in the figure.
One can notice that 
the curve for $\Lambda=2\gamma$ is insensitive to $g(\Lambda)$,
except for a singular point which corresponds to a three-body bound state
with zero binding energy, $B_3=0$,
even though the value of $g(\Lambda)$ is significantly changed.
We note that the $S$-wave three-body force for the spin quartet channel
may be suppressed 
due to the Pauli principle (by applying the antisymmetrization 
operator) in the conventional potential model calculation.
We can reproduce the same effect (except for the appearance of the 
bound state) as that of the Pauli principle when we solve the integral
equation using the normal cutoff value.
The curves with 
$\Lambda=\gamma$ and $0.5\gamma$, on the other hand,
 show a sensitivity to $g(\Lambda)$ 
and vary widely and smoothly.

In the right panel of Fig.~\ref{fig;a4Lam30}, 
we numerically calculate $g(\Lambda)$ from Eq.~(\ref{eq;a}) 
and plot curves of $g(\Lambda)$ 
which reproduce $a_4^{th}$
(curve) and the three-body bound state with 
$B_3=0$ (dashed curve) 
as functions of $\Lambda$. 
One can see that when the cutoff value is large, $\Lambda > 160$~MeV,
the value of $g(\Lambda)$ almost vanishes. 
This may indicate the effect of the Pauli principle for
the spin quartet channel.
As the cutoff value is further reduced 
smaller than $\Lambda = 160$~MeV, we need nonzero value of $g(\Lambda)$ 
where the short range length scale of the theory becomes 
$r (=\Lambda^{-1}) >1.24$~fm.
This length scale might be regarded long enough 
to be out of the range of the Pauli principle.
We note that the non-vanishing three-body contact interaction 
we obtained here
may not be a genuine one, but correspond to the one 
induced by
the exchanging nucleon propagator of the larger momentum than the value of 
$\Lambda$, which connects two two-body interactions and 
makes the effective three-body one at the small cutoff values.  
The similar observation that a three-body force is generated 
from two-body forces at small cutoffs in the SRG analysis is 
reported in Ref.~\cite{jnf-prl09}.
%
One can also see that the three-body bound state
with $B_3=0$ appears when 
the strength of $g(\Lambda)$ becomes stronger
than that of $g(\Lambda)$ which reproduces $a_4^{th}$
and $\Lambda$ is larger than about 60~MeV in the 
figure. 
We find that the curve of $g(\Lambda)$ for the three-body bound 
state with $B_3=0$ varies smoothly, whereas that of $g(\Lambda)$ which
reproduces $a_4^{th}$ has a plateau like shape at the middle of 
the range of $\Lambda$, 
$\Lambda \simeq 60 \sim$ 100~MeV.
We use the curve of $g(\Lambda)$ which reproduces $a_4^{th}$ 
when studying the perturbation expansions, 
the ordinary and inverted Born series, 
in the following.

\section{Perturbative expansions: the ordinary and inverted Born series}
\label{sec:3}

Now we expand the scattering length and the amplitude 
in terms of the ordinary and inverted Born series, 
as discussed in the introduction.

\subsection{The ordinary and inverted Born series for the scattering length}
\label{sec:3.1}

Firstly, we study the expansion in terms of 
the ordinary and inverted Born series for
the scattering length $a_4$.
Thus the scattering length $a(0,0)$ in Eq.~(\ref{eq;a}) is expanded as
\bea
a_4 &=& a(0,0) = 
\frac{8Z_d}{3\rho_d}\left[
b_0 + b_1 + b_2 + \cdots\right] \,,
\label{eq;a-born1}
\eea
where 
\bea
b_0 &=& \frac{1}{\gamma^2} + g(\Lambda)\,,
\\ 
b_1 &=& 
\int^\Lambda_0 dq\left[
\frac{1}{\gamma^2+q^2} +g(\Lambda)\right]^2
F(q)\,,
\\
b_2 &=& 
\int^\Lambda_0 dq\left[
\frac{1}{\gamma^2+q^2}+ g(\Lambda)\right]F(q)
\nnb \\ && \times
\int^\Lambda_0 dq'\left[
K^{(a)}(q,q',-B_2)+g(\Lambda)\right]F(q')\left[
\frac{1}{\gamma^2+q'^2}+g(\Lambda)\right]\,,
\eea
with
\bea
F(q) &=& 
- \frac{8}{3\pi}
\frac{\gamma + \sqrt{\gamma^2 + \frac34q^2}}{
1 - \frac12\rho_d \left(
\gamma
+ \sqrt{\gamma^2 + \frac34q^2}
\right)}\,.
\eea 
This expansion corresponds to that around the weak coupling limit.
On the other hand, we consider another expansion,
the inverted Born series, of the scattering length as
\bea
\frac{1}{a_4} &=& 
\frac{3\rho_d}{8Z_d}\left[
\frac{1}{b_0}
-\frac{b_1}{b_0^2}
-\frac{b_2}{b_0^2}
+\frac{b_1^2}{b_0^3}
+ \cdots
\right]\,.
\label{eq;a-born2}
\eea

\begin{figure}[t]
\begin{center}
\includegraphics[width=7cm]{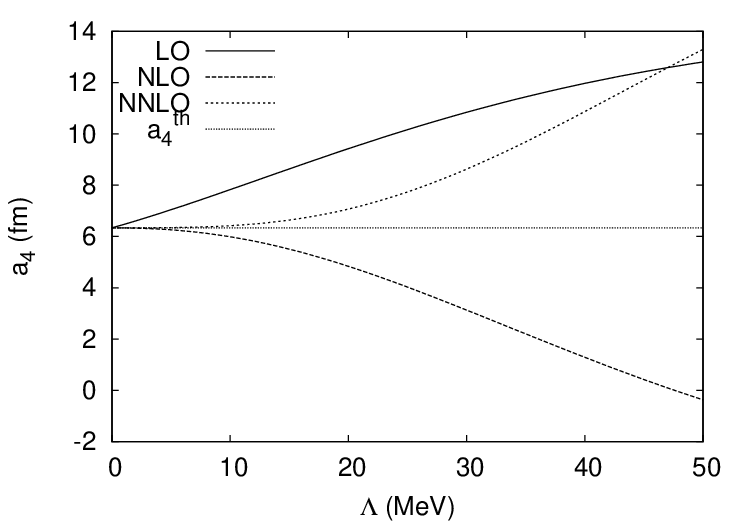}
\includegraphics[width=7cm]{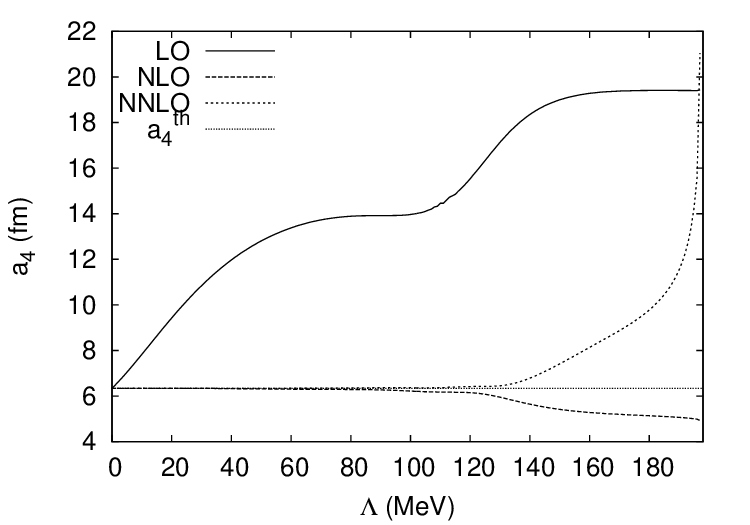}
\caption{Left panel: 
Scattering length $a_4$ 
obtained from the ordinary Born expansion
as functions of $\Lambda$~(MeV). 
Curves labeled by ``LO'' are results at leading order, 
``NLO'' up to next-to-leading order,
and ``NNLO'' up to next-to-next-to leading order
in the both panels. 
A horizontal line of $a_4^{th}$ is also included. 
Right panel:
Scattering length $a_4$ 
obtained from the inverted Born expansion
as functions of $\Lambda$~(MeV). 
}
\label{fig;a4-born-Lam12}
\end{center}
\end{figure}

In the left and right panel of Fig.~\ref{fig;a4-born-Lam12}, 
we numerically calculate and plot 
curves of the scattering length $a_4$ 
obtained from the ordinary and inverted Born series,
respectively,
as functions of $\Lambda$.  
Curves labeled by ``LO" are 
results which include only 
the leading order term, $b_0$,  
those by ``NLO" are results up to next-to-leading order (NLO)
which include first two terms in the brackets
in Eq.~(\ref{eq;a-born1}) and (\ref{eq;a-born2}), 
and those by ``NNLO" are results up to next-to-next-to leading
order (NNLO) which include all terms  in the brackets.

In the left panel of Fig.~\ref{fig;a4-born-Lam12}, 
one can see that a region of the cutoff $\Lambda$,
where the curves of $a_4$ obtained from the terms up to NLO and NNLO 
in the ordinary Born expansion
agree with $a_4^{th}$, is quite small, up to about 10~MeV. 
It would be a natural consequence of 
the perturbation around the 
weak coupling limit because the perturbation would converge 
when the scale $\Lambda$ becomes smaller than a typical scale 
of the process.  In the present case, it may be $a_4^{th}$ 
($1/a_4^{th}\simeq 31.4$~MeV), and thus it converges when
the cutoff $\Lambda$ becomes much smaller than $1/a_4^{th}$. 
In the right panel of Fig.~\ref{fig;a4-born-Lam12}, 
on the other hand, we find that the region where $a_4^{th}$
is reproduced is remarkably broaden,  up to about 100~MeV
due to the two or three terms (up to NLO and NNLO, respectively)
of the inverted Born series.

\subsection{The ordinary and inverted Born series for the scattering amplitude}
\label{sec:3.2}

Before expanding the integral equation 
for the scattering amplitude
in terms of the ordinary and inverted Born series, 
we check how the three-body interaction $g(\Lambda)$ 
can reproduce the phase shift $\delta_0$ 
at the small cutoff values. 
\begin{figure}[h]
\begin{center}
\includegraphics[width=10cm]{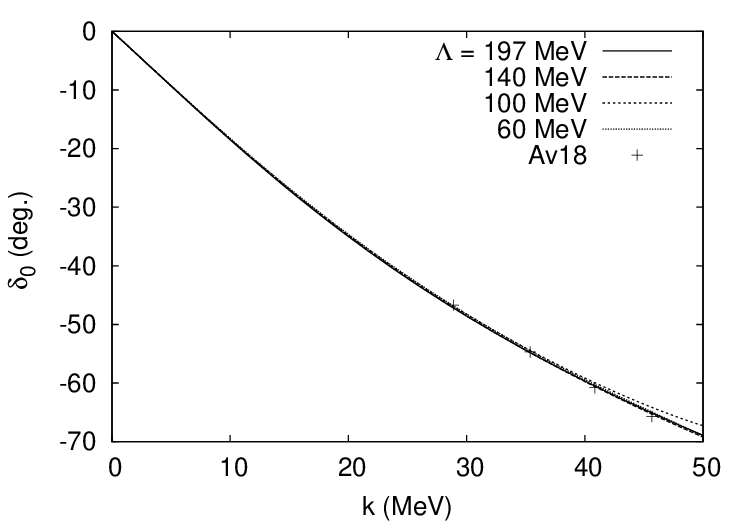}
\caption{
Phase shift $\delta_0$ (deg.) 
of the $S$-wave $nd$ scattering for spin quartet channel
below deuteron breakup threshold
as functions of $k$ (MeV).
Curves are obtained by solving the integral equation 
at cutoff values, $\Lambda =197$, 140, 100, and 60~MeV with the 
three-body interaction $g(\Lambda)$ renormalized by $a_4^{th}$.  
Plus signs "+" denote results from a modern potential 
model (Av18)~\cite{ketal-npa96}. 
The deuteron breakup momentum is $k_{br}\simeq 52.7$~MeV.
}
\label{fig;del0-full}
\end{center}
\end{figure}
In Fig.~\ref{fig;del0-full}, we numerically calculate and plot the 
phase shift $\delta_0$  below deuteron breakup threshold
as functions of the on-shell momentum $k$
by solving the integral equation 
at small cutoff values, $\Lambda =197$, 140, 100, and 60~MeV.
The deuteron breakup momentum is $k_{br}\simeq 52.7$~MeV. 
We also include results from a modern 
potential model (Av18)~\cite{ketal-npa96} in the figure.
One can see that the results are 
fairly independent on the cutoff
values and considerably well agree to those obtained 
from the accurate Av18 potential model.

We now expand the on-shell T-matrix in terms of 
the ordinary and inverted Born series up to NNLO.
Thus the on-shell T-matrix is obtained as 
\bea
T(k,k) &=& - \frac{8\pi Z_d}{m_N\rho_d}\left[
B_0 + B_1 + B_2 + \cdots
\right]\,,
\label{eq;on-T1}
\eea 
with
\bea
B_0 &=& V(k,k) = K^{(a)}(k,k;E) + g(\Lambda)\,,
\\
B_1 &=& 
\int^\Lambda_0 dq\, V(k,q) G(q,k)V(q,k)\,,
\\
B_2 &=& \int^\Lambda_0 dq V(k,q)G(q,k)\ 
\int^\Lambda_0 dq'\, V(q,q') G(q',k) V(q',k)\,,
\eea
where
\bea
G(q,k) &=& - \frac{8}{3\pi}
\frac{\gamma + \sqrt{\gamma^2 + \frac34(q^2-k^2)}}{
1-\frac12\rho_d\left[\gamma+ \sqrt{\gamma^2+\frac34(q^2-k^2)}\right]}
\frac{q^2}{q^2-k^2-i\epsilon}\,.
\eea
This corresponds to the expansion around the weak coupling limit.
We also have an expansion around the inverse of the T-matrix as
\bea
\frac{1}{T(k,k)} &=& - \frac{m_N\rho_d}{8\pi Z_d}\left[
\frac{1}{B_0}
- \frac{B_1}{B_0^2}
- \frac{B_2}{B_0^2}
+ \frac{B_1^2}{B_0^3}
+ \cdots
\right]\,.
\label{eq;on-T2}
\eea

To calculate the phase shift $\delta_0$ from 
the ordinary Born series 
in Eq.~(\ref{eq;on-T1})
we employ the relation
\bea
\delta_0 &=& \frac{1}{2i}\ln\left[
1 
+ i \frac{2km_N}{3\pi}T(k,k)
\right]\,,
\eea
We note that 
when the amplitude expanded in the ordinary Born series is truncated, 
it breaks unitary 
condition and  the phase shift becomes a complex number~\cite{ah-prc12}.
On the other hand, to calculate the phase shift from the 
inverted Born expansion 
in Eq.~(\ref{eq;on-T2})
we use the formula 
\bea
k\cot\delta_0 = \frac{3\pi}{m_N} Re\frac{1}{T(k,k)}\,,
\eea
with
\bea
Re\frac{1}{T_{NLO}(k,k)} &=& 
- \frac{m_N\rho_d}{8\pi Z_d}\left[
\frac{1}{B_0} 
- \frac{Re B_1}{B_0}
\right]\,,
\label{eq;1/T-NLO}
\\
Re\frac{1}{T_{NNLO}(k,k)} &=& 
- \frac{m_N\rho_d}{8\pi Z_d}\left[
\frac{1}{B_0} 
- \frac{Re B_1}{B_0^2}
- \frac{Re B_2}{B_0^2}
+\frac{(ReB_1)^2-(Im B_1)^2}{B_0^3}
\right]\,,
\label{eq;1/T-NNLO}
\eea
where $B_0$ has real part only.
This expansion preserves the unitary condition,
at least up to NNLO.

\begin{figure}[t]
\begin{center}
\includegraphics[width=10cm]{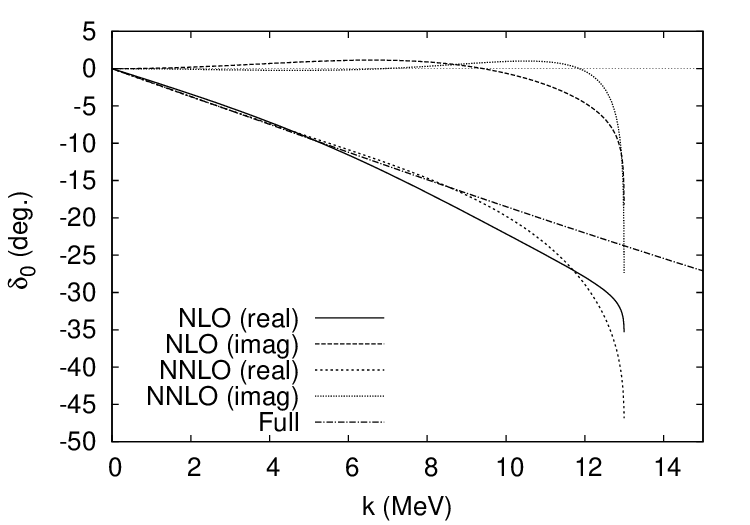}
\caption{Real and imaginary part of phase shift
$\delta_0$ (deg.) of $S$-wave spin quartet $nd$ scattering, 
as functions of $k$ (MeV), obtained from the 
truncated ordinary Born series up to NLO and NNLO
where the cutoff is fixed at $\Lambda = 13$~MeV.
A dashed line labeled by ``Full'' obtained from the calculation 
without the truncation and $\Lambda=197$~MeV
is also included.
}
\label{fig;del0-born-1}
\end{center}
\end{figure}

In Fig.~\ref{fig;del0-born-1},
we numerically calculate and plot 
real and imaginary part of the phase shift $\delta_0$ 
of $S$-wave spin quartet $nd$ scattering 
obtained from the truncated ordinary Born series 
in Eq.~(\ref{eq;on-T1}). 
Curves up to NLO are obtained by including first two terms,
$B_0$ and $B_1$, and those up to NNLO
by including all three terms, $B_0$, $B_1$, and $B_2$, 
in Eq.~(\ref{eq;on-T1}). 
We have fixed $\Lambda$ at $\Lambda = 13$~MeV 
because we found 
in the left panel of Fig.~\ref{fig;a4-born-Lam12}
that the scattering length 
$a_4^{th}$ are fairly well reproduced when the cutoff value is about up 
to 10~MeV. 
A line labeled by ``Full''\footnote{
This line is the same as that at $\Lambda=197$~MeV 
in Fig.~\ref{fig;del0-full}.}
obtained from the calculation without the truncation at $\Lambda=197$~MeV
is also included in the figure.  One can see that the real parts of 
$\delta_0$ agree with 
the full result up to about 6~MeV for NLO and 9~MeV for NNLO.  
In addition, the imaginary parts sharply decrease 
around the edge of the cutoff value $\Lambda= 13$~MeV, 
and it indicates that the unitary condition is broken.
The broken unitary condition up to NLO is also partly cured 
by including the higher order term, $B_2$, at NNLO.

\begin{figure}[t]
\begin{center}
\includegraphics[width=10cm]{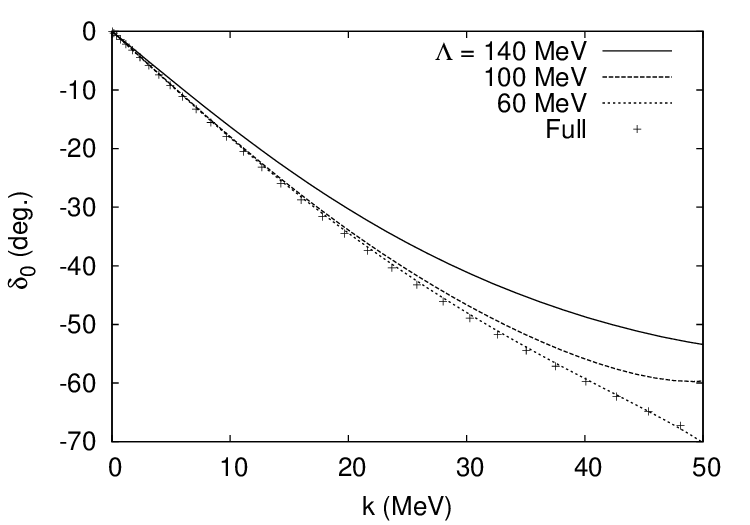}
\caption{
Phase shift $\delta_0$ (deg.) of $S$-wave spin quartet $nd$ scattering, 
as functions of $k$ (MeV), obtained from 
the inverted Born series up to NLO 
with $\Lambda=140$, 100, and 60~MeV.
Plus signs ``+'' labeled by ``Full'' 
are obtained from the calculation without
the truncation at $\Lambda=197$~MeV.
}
\label{fig;del0-born-2-nlo}
\end{center}
\end{figure}

\begin{figure}[t]
\begin{center}
\includegraphics[width=10cm]{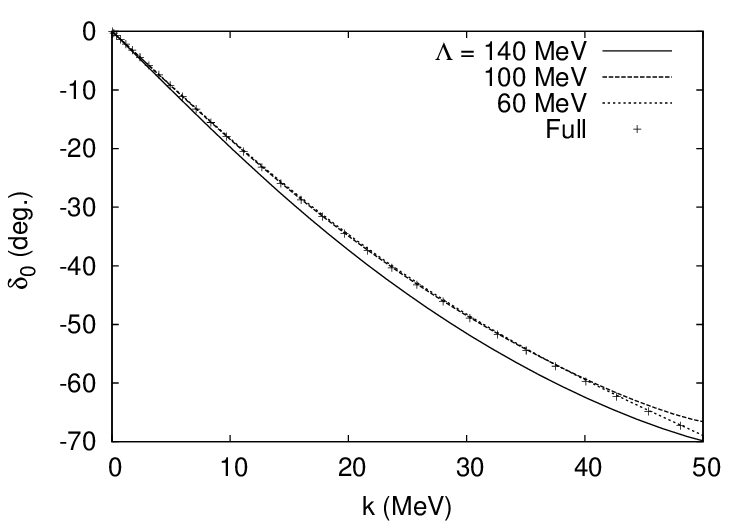}
\caption{
Phase shift $\delta_0$ (deg.) 
obtained from the inverted Born series up to NNLO
with $\Lambda=140$, 100, and 60~MeV.
See the caption of Fig.~\ref{fig;del0-born-2-nlo} as well.
}
\label{fig;del0-born-2-nnlo}
\end{center}
\end{figure}

In Figs.~\ref{fig;del0-born-2-nlo} and \ref{fig;del0-born-2-nnlo},
we numerically calculate and plot curves of the phase shift $\delta_0$ 
obtained from the inverted Born series up to NLO 
and NNLO in Eqs.~(\ref{eq;1/T-NLO}) and (\ref{eq;1/T-NNLO}), 
respectively, 
where the cutoff values are chosen $\Lambda=140$, 100, and 60~MeV. 
In addition, plus signs ``+'' labeled by ``Full''  in the figures 
denote the results from the calculation without 
the truncation at $\Lambda = 197$~MeV. 
We find in Fig.~\ref{fig;del0-born-2-nlo} that 
the curves considerably well converge to that of the full calculation
as the cutoff value decreases, and 
the truncated inverted Born series 
up to NLO fairly well reproduces the result of the 
full calculation at $\Lambda=60$~MeV. 
In Fig.~\ref{fig;del0-born-2-nnlo}, we can see that 
the convergence to the full result, 
as reducing the cutoff value, becomes faster
due to the inclusion of the higher order terms.

\section{Discussion and conclusions}
\label{sec:4}

In this work, we studied the perturbative expansions,
as the ordinary and inverted Born series, of the integral equation 
for the $S$-wave $nd$ scattering for 
the spin quartet channel below the deuteron breakup 
threshold in pionless EFT at the small cutoff values.
The three-body contact interaction is introduced 
when the integral equation is solved by using the small 
cutoff values. 
After the strength of the three-body interaction is 
renormalized by using the scattering length $a_4$, 
we expand the integral equation for the scattering 
length and the amplitude, as the ordinary and inverted Born series, 
up to NNLO. 
We find that the scattering length (the phase shift)
is considerably well reproduced by a few terms of the 
ordinary and inverted Born series as we reduce the cutoff values 
to about 10~MeV (10~MeV)
and to about 100~MeV (60~MeV), respectively.
Therefore, the inverted Born expansion in the present process
can be a relevant approximation with a significantly larger valid momentum
than the ordinary Born expansion.

In the present particular process, 
the $S$-wave spin quartet $nd$ scattering, 
one may regard that there is no advantage 
by sending $\Lambda$ to small values 
because the physical observables, the scattering 
length $a_4$ and the phase shift $\delta_0$, are well described
(without fitting any unknown parameters) 
by the two effective range parameters in the deuteron 
channel at the usual large scale of the pionless theory,
$\Lambda\simeq m_\pi$.   
However, it could be a useful limit 
when one studies, e.g., a relation between pionless EFT and 
a Halo/Cluster EFT whose large scale smaller than $m_\pi$,
such as a deuteron cluster theory 
for a reaction whose typical scale $Q$ is smaller than 
the deuteron breakup momentum~\cite{sia}. 

In addition,
it may be the interesting observation that 
the physical observables can be well described by a few terms of the 
inverted Born series, 
being closely related to  
the unitary limit, 
at the relatively large cutoff values. 
If this property were common in some class of the interactions and/or in 
the unitary limit,
it could provide us a useful method to make 
a non-perturbative interaction perturbative 
and/or to be used in studies of the exotic nuclei near the drip line.
Now we are studying the property of the inverted Born expansion
by employing a renormalization group analysis,
and it is to be reported separately.

\begin{acknowledgements}
The author would like to thank K. Kubodera for reading the manuscript
and Y.-H. Song for discussion.
This work is supported by the Basic Science Research Program through
the National Research Foundation of Korea (NRF) funded by the Ministry 
of Education, Science and Technology (2010-0023661) and 
(2012R1A1A2009430).
\end{acknowledgements}

\end{document}